\documentclass[a4paper,10pt]{article}

\usepackage{graphicx}
\usepackage{amsmath}

\newcommand{\la}{\label}
\newcommand{\p}{\partial}
\newcommand{\be}{\begin{equation}}
\newcommand{\ee}{\end{equation}}
\newtheorem{thm}{Theorem}[section]
\newtheorem{definition}{Definition}[section]
\newcommand{\ba}{\begin{eqnarray}}
\newcommand{\ea}{\end{eqnarray}}

\newcommand{\kk}{\kappa}

\begin{document}\centerline{\bf \begin{Large} The Penrose inequality for nonmaximal perturbations \end{Large}} 

\centerline{\bf \begin{Large} of the Schwarzschild initial data \end{Large}}                                                                                    

\bigskip

\centerline{J. Kopi\'nski and J. Tafel}

\centerline{Faculty of Physics, University of Warsaw,}
\centerline{Pasteura 5, 02-093 Warsaw, Poland}
\begin{abstract}
We show  that the Penrose inequality  is satisfied for a class of conformally flat axially symmetric nonmaximal perturbations of the Schwarzschild data. A role of horizon is played by a marginally outer trapped surface which does not have to be minimal.
\end{abstract}

\null
\section{Introduction}
The Penrose-Hawking \cite{p1,h1} theorems predict a development of singularities in solutions of the Einstein  equations if a marginally trapped surface is admitted. According to the cosmic censorship conjecture of Penrose \cite{p2} these singularities  should be hidden behind event horizons of black holes. A heuristic argument supporting this statement relied on a consideration of collapsing null shells and led to the inequality between the ADM mass $M$ and the area of an apparent horizon $|S_h|$
\be\la{1a}
M\geq \sqrt{\frac{|S_h|}{16\pi}},
\ee
called the Penrose inequality. Currently, it is expected that (\ref{1a}) should hold for any initial surface $S$ with an inner boundary $S_h$ which   represents surface of a black hole. The Penrose inequality was first proved for the generic apparent horizon in the case of spherical symmetry \cite{mn}. A mathematically complete proof for a wide class of data was given by  Huisken and Ilmanen \cite{hi}. They established the so-called Riemannian Penrose inequality in the case of  horizon represented by a minimal surface. This approach utilized a scenario, proposed by Jang and Wald \cite{jw}, based on the monotonicity of the Geroch quasi-local mass  \cite{g} under the inverse mean curvature flow (IMCF).  A generalization to multi-connected horizons was found by   Bray \cite{br}. Later some analytical methods have been proposed to prove (\ref{1a}) in general case, but  technical complexity of these approaches prevents them from being conclusive \cite{mms,bk}. 

In the case of axially symmetric data the inequality (\ref{1a}) can be replaced by its stronger version,
\be\la{8k}
E^2-p^2\geq \frac{|S_h|}{16\pi}+\frac{4 \pi}{|S_h|}J^2,
\ee
where $E$, $p$  and $J$ are, respectively, energy, momentum and angular momentum of initial data.
In paper \cite{kt} we studied inequality (\ref{8k}) in the case of axially symmetric conformally flat maximal data on $R^3$ with removed ball of radius $m/2$. The internal boundary was a marginally outer trapped surface (MOTS). Assuming that data can be expanded with respect to a small parameter $\epsilon$ we proved (\ref{8k}) up to leading terms in $\epsilon$ (typically up to $\epsilon^2$). This means that in the considered class any initial data in a  neighbourhood of the Schwarzschild data satisfy the Penrose inequality. 

In this paper we continue our approach from \cite{kt} but now we admit nonvanishig mean curvature of data (nonmaximal data). This change makes constraint equations more complicated and adds a new free function in data. Using a slightly new method we are still able to prove (\ref{8k}) up to $\epsilon^2$ in a generic case.

The main value of this result lies in a relaxation of typical assumptions in almost all other proofs ($H=0$ and a minimal surface as a MOTS, see \cite{rm} for an exception). The main disadvantage of our approach is that the considered class of data is relatively small (3 free functions of 2 coordinates). Moreover,  we don't know if the inner MOTS is outermost (see a discussion in Section 1 in \cite{kt}) and we are not able to treat nongeneric case (see Definition 3.1).

The remainder of this paper is organized as follows. In section 2 we describe shortly the conformal approach to constraint equations with a boundary condition which guarantees that the internal boundary is MOTS. We also perform expansions in $\epsilon$ and give  a perturbative formula for an expression $P_I$ which is crucial for the Penrose inequality. In section 3 we find axially symmetric solutions of  the momentum constraint in the leading order in $\epsilon$ and estimate function $P_I$ from below. The main result is formulated in section 4.

\section{A perturbative formulation of the Penrose inequality} 
Initial data $g'_{ij},K'_{ij}$ on a surface $S$ are constrained by equations
\begin{align}
\nabla_i\big(K'^i{}_{j}-H'\delta^i_{\ j}\big)&=0\ ,\la{1x}\\
  R'+H'^2-K'^2&=0\ ,\la{1y}
\end{align}  
where $H'=K'^i{}_{ i}$, $K'^2=K'_{ij}K'^{ij}$ and $R'$ is the Ricci scalar of $g'_{ij}$.
We are interested in  data which are asymptotically flat and admit a marginally outer trapped surface (MOTS) $S_h$ with  vanishing expansion of outer null rays,
\be\la{2x}
H'-K'_{nn}+ h=0\ \ \mathrm{on}\ \ S_h\ ,
\ee
where $K'_{nn}=n'^in'^jK'{}_{ij}$, $n'^i$ is the outer  unit normal vector  and $h=\nabla_in'^i$ is the mean curvature of the surface.

In the conformal approach  initial data are parametrized in the following way \cite{y}  by a preliminary metric $g_{ij}$, symmetric traceless tensor $A_{ij}$ and the mean curvature $H'$,
\be\la{3x}
g'_{ij}=\psi^4 g_{ij},\quad K'^i{}_{ j}=\psi^{-6}A^i_{\ j}+\frac 13H'\delta^i_{\ j}\ ,\ \ A^i_{\ i}=0\ .
\ee
For nonmaximal data ($H'\neq 0$) constraints
(\ref{1x})-(\ref{1y})   form a coupled system of equations
\be\la{4x}
\nabla_iA^i_{\ j}=\frac 23\psi^6\nabla_jH'\ ,
\ee
\begin{equation}\la{5x}
\bigtriangleup\psi=\frac 18 R\psi-\frac 18  A_{ij}A^{ij}\psi^{-7}+\frac {1}{12}H'^2\psi^5\ ,
\end{equation}
and the boundary condition (\ref{2x}) is replaced by
\be\la{6x}
n^i\p_i \psi+\frac 12 h\psi-\frac 14 A_{nn} \psi^{-3}+\frac 16H'\psi^3=0\ \ \mathrm{on}\ \  S_h.
\ee
Unfortunately, in the asymptotically flat setting there is no existence theorems for the system (\ref{4x})-(\ref{6x}). We suppose that results of Holst and Meier \cite{ho1} could be adapted to this case  (note that in Theorem 3.2 in \cite{ho1} it is assumed that $\theta_-=0$ in contrary to our assumption $\theta_+=0$ represented by (\ref{2x})). A review of results on existence theorems in different settings can be found in \cite{dhkm}.

As in \cite{kt},   we assume in this paper that $g_{ij}=\delta_{ij}$ and the initial surface is given by  $S=R^3\backslash B(0,\frac m2)$, where  $B(0,\frac m2)$ is an open ball  with radius $m/2$ and the spherical boundary  $S_h$. 
If $K'_{ij}=0$  then solution of (\ref{5x})-(\ref{6x}) reads
\be\la{9}\psi=\psi_0=1+\frac{m}{2r}
\ee
and transformation (\ref{3x}) leads  to the Schwarzschild initial metric,
\be 
g'_0=\frac{dr'^2}{1-\frac{2m}{r'}}+r'^2(d\theta^2+\sin^2{\theta}d\varphi^2)\ ,
\ee
for which the Penrose inequality is saturated. We are going to investigate  this inequality for  data with small addition of $A_{ij}$ and nontrivial $H'$.

 Unlike in the case $H'=0$, equations (\ref{4x}) cannot be solved independently of (\ref{5x}). However, this can be done approximately under the assumption that $\psi$,  $A_{ij}$ and $H'$ can be expanded into powers of a small parameter $\epsilon$,
\be\la{7x}
\psi=\psi_0+\psi_1+\psi_2+...
\ee
\be\la{8x}
A^{ij}=A_1^{ij}+A_2^{ij}+...\ \ \ H'=H'_1+H'_2+...
\ee
Here, subscripts n=0,1,2.. correspond to terms of the order  $\epsilon^n$. We will denote the leading terms in (\ref{8x}) by $B_{ij}$ and $B$,
\be\la{9x}
B^{ij}=A_1^{ij}\ ,\ \ B=H'_1\ .
\ee
In the lowest nontrivial order in $\epsilon$ the  constraints read
\be\la{10x}
\nabla_iB^i_{\ j}=\frac 23\psi_0^6\nabla_jB\ ,
\ee
\begin{equation}\la{11x}
\bigtriangleup\psi=-\frac 18  B_{ij}B^{ij}\psi_0^{-7}+\frac {1}{12}B^2\psi_0^5\ .
\end{equation}
Thus, one can first solve  (\ref{10x}) and then extract information about total energy and area of the MOTS from equation (\ref{11x}). Considerations from Section 2 in \cite{kt} can be almost literally repeated  under  the asymptotical flatness conditions of the form
\be\la{10y}
(B_{ij},B)=O(r^{-2})\ ,\ \ \p_p(B_{ij},B)=O(r^{-3})\ ,\ \ \psi\longrightarrow 1\ \ \mathrm{if}\ \ r\longrightarrow\infty.
\ee
Up to $\epsilon^2$ one obtains
\be\la{10z}
E^2-\frac{|S_h|}{16\pi}=P_I\ ,
\ee
where
\be\la{10u}
P_I=(\frac{m^2}{128\pi})^2\langle B_{rr}-\frac 23B\psi_0^6\rangle_{h}^2-\frac{3m^2}{8\pi}\langle \psi_1^2\rangle_{h}+\frac{m}{8\pi}\int_{\frac m2}^{\infty}r^2(2-\psi_0)\langle B_{ij}B^{ij}\psi_0^{-7}-\frac 23B^2\psi_0^{5}\rangle dr\ .
\ee
Here $\langle\rangle$ denotes integral over the 2-dimensional sphere with radius $r$ and the subscript $h$ refers to the sphere $r=m/2$.
The function $\psi_1$ is a solution of 
the flat Laplace equation,
\be\la{14a}
\bigtriangleup\psi_1=0\ ,
\end{equation}
with the boundary conditions
\be\label{15}
\p_r\psi_1+\frac 1m\psi_1=\frac {1}{32} (B_{rr}-\frac 23B\psi_0^6)\ \ \mathrm{at}\ \ r=m/2\ ,
\ee
\be\la{16}
\psi_1= 0\ \ \mathrm{at}\ \ r=\infty\ .
\ee
The standard Penrose inequality is satisfied if $P_I\geq p^2$. In the next section we  consider all axially symmetric solutions of (\ref{10x}) and we show that in a generic case $P_I\geq p^2+\frac{J^2}{4m^2}$, hence inequality (\ref{8k}) is satisfied up to $\epsilon^2$.
\section{Axially symmetric perturbations}
Equation  (\ref{10x}) with index $\varphi$ yields
\begin{equation}\la{12x}
B_{\varphi\theta}=\frac{\omega _{,r}}{\sin{\theta}}\ ,\ \ B_{\varphi r}=-\frac{\omega _{,\theta}}{r^2\sin{\theta}}\ , 
\end{equation}
with an arbitrary function $\omega$, 
and  for other indices we obtain the following system,
\be\la{13x}
(r^3B_{rr}\sin{\theta})_{,r}+(rB_{r\theta}\sin{\theta})_{,\theta}=\frac 23r^3\psi_0^6B_{,r}\sin{\theta}\ ,
\ee
\be\label{14x}
(B_{\theta\theta}\sin^2{\theta})_{,\theta}+(r^2B_{r\theta})_{,r}\sin^2{\theta}+r^2B_{rr}\sin{\theta}\cos{\theta}=\frac 23r^2\psi_0^6B_{,\theta}\sin^2{\theta}\ .
\ee
In order to solve equations (\ref{13x})-(\ref{14x}) let us introduce function $S$ such that
\be\la{3}
B=\frac 32S_{,z}\ ,
\ee
where $z=\cos{\theta}$. 
Equation (\ref{13x}) leads to the existence of another function  $Q$ such that
\be\label{4}
B_{r\theta}=\frac{1}{r\sin{\theta}}(Q_{,r}-\kk S_{,r})\  ,
\ee
\be\label{6}
B_{rr}=\frac{1}{r^3}Q_{,z}\ ,
\ee
where
\be\la{7}
\kappa =r^3\psi_0^6\ .
\ee

Following \cite{kt}, we also introduce function $F(r,z)$ according to
\be\la{11}
B_{\theta\theta}+\frac 12r^2B_{rr}=F_{,z}\ .
\ee
Now equation (\ref{14x}) takes the form
\be\label{12}
\bigtriangleup_s F=(rQ_{,r}-r\kappa S_{,r})_{,r}-\frac {1}{2r} (z^2-1)(Q+2\kappa S)_{,zz}\ ,
\ee
where
\be
\bigtriangleup_s F=((1-z^2)F_{,z})_{,z}\ .
\ee
 Function $F$ can be easily found if the r.h.s. of equation (\ref{12})  is expressed in terms of the Legendre polynomials $P_n$. A  necessary condition  is the absence of  $P_0$  in (\ref{12}), hence
\be\label{12a}
(rQ_{0,r}-r\kappa S_{0,r})_{,r}+\frac {1}{3r}((Q_{,z})_1+2\kappa (S_{,z})_1)=0\ ,
\ee
where subscripts $0,1$ denote respective coefficients of  decompositions into the Legendre polynomials.

Potentials $S,Q$  define smooth quantities $B_{ij}$ and $B$, provided they are smooth  functions of $r,z$ and 
\be\la{16}
(Q_{,r}-\kk S_{,r})\sim \sin^2{\theta}\ .
\ee
The asymptotic flatness conditions (\ref{10y}) yield
\be\la{12b}
r^2S,\frac 1rQ<\infty\ .
\ee

Function (\ref{10u}) can be written in the form
\ba\la{32g}
P_I=(\frac{m^2}{128\pi})^2\langle B_{rr}-\frac 23B\psi_0^6\rangle_{h}^2-\frac{3m^2}{8\pi}\langle \psi_1^2\rangle_{h}
+P_J+\tilde P_I\ ,
\ea
where
\be\la{19a}
P_J=2\int_{\frac m2}^{\infty}{dr \varrho r\int_{-1}^1(B_{\varphi\theta}^2+r^2B_{\varphi r}^2 )\frac{dz}{1-z^2}}\ ,
\ee
\be\la{19}
\tilde P_I=\int_{\frac m2}^{\infty}dr \varrho r\int_{-1}^1I dz
\ee
and
\be\la{18}
\varrho=\frac {m(1-\frac {m}{2r})}{4r^3\psi_0^7}\ ,
\ee
\be\la{19a}
I=2r^2(B_{r\theta})^2+2(B_{\theta\theta}+\frac 12r^2B_{rr})^2+\frac 32r^4(B_{rr})^2-\frac23r^4 \psi_0^{12}(B)^2\ .
\ee
Because of (\ref{11}) and  (\ref{12}) $I$ contains squares of second derivatives with respect to $r$. This makes estimating  integral (\ref{19}) rather difficult. In order to reduce order of derivatives in $I$ we replace  $Q,S$ by new variables $U,W$ given by
\be\la{13}
U=Q_{,r}-\kk S_{,r}\ ,\ \ W=Q_{,z}-\kk S_{,z}\ .
\ee
In terms of $U,W$ equations (\ref{3})-(\ref{6}) and (\ref{12}) take the form
\be\la{8}
B=\frac{3}{2\kk_{,r}}(U_{,z}-W_{,r})\  ,
\ee
\be\label{8a}
B_{r\theta}=\frac{U}{r\sin{\theta}}\  ,
\ee
\be\label{9}
B_{rr}=\frac{1}{r^3}(W+\frac{\kk}{\kk_{,r}}(U_{,z}-W_{,r}))\ ,
\ee
\be\label{9a}
\bigtriangleup_s F=(rU)_{,r}-\frac {1}{2r} (z^2-1)(W+\frac{3\kk}{\kk_{,r}}(U_{,z}-W_{,r}))_{,z}\ .
\ee

Since 
$$\kk_{,r}=3\alpha r^2\psi_0^5\ ,$$
where
\be\la{9b}
\alpha=1-\frac {m}{2r}\ ,
\ee
functions  $B$ and $B_{rr}$ are finite at $r=m/2$ iff 
\be\la{15}
U_{,z}=W_{,r}\ \ \mathrm{at}\ \ r=\frac m2\ .
\ee
Conditions (\ref{12a})-(\ref{16}) transform into
\be\label{22}
r(rU_0)_{,r}+\frac{1}{3}W_1+\frac{\kk}{\kk_{,r}}((U_{,z})_1-W_{1,r})=0\ ,
\ee
\be\la{24}
U,\frac 1r W<\infty\ ,
\ee
\be\la{25}
U\sim (z^2-1)\ .
\ee

In terms of $U,W$ function (\ref{19a})  reads
\be\la{20}
I=\frac{2U^2}{1-z^2}+2F_{,z}^{\ 2}+\frac{3}{2r^2}W(W+\frac{2\kk}{\kk_{,r}}(U_{,z}-W_{,r}))\ .
\ee
Substituting (\ref{20}) to (\ref{19}) and integrating by parts  the  term proportional to $W_{,r}$ in (\ref{20}) leads to
\be\la{23}
\tilde P_I=\int_{\frac m2}^{\infty}{dr \rho r\int_{-1}^1 \tilde I dz}+\frac{\pi}{8m^3}\int_{-1}^1{(W^{2})_hdz}\ ,
\ee
where index $h$ denotes value at $r=m/2$ and 
\be\la{27}
\tilde I=\frac{2U^2}{1-z^2}+2F_{,z}^{\ 2}+\frac{\psi_0}{r\alpha }WU_{,z}\ .
\ee

In order to satisfy condition (\ref{25}) and  assure simplicity of (\ref{23}) if the dependent variables are  expanded into the Legendre polynomials it is convenient to define new variables $X,Y$ by
\be\la{26}
U=\psi_0^3(z^2-1)Y_{,z}\ ,\ \ W=2r\psi_0^2(X-\alpha Y)\ .
\ee
In accordance with the regularity conditions they are smooth functions of $r$ and $z$, bounded at infinity. Condition (\ref{15}) yields
\be\la{27}
Y-\bigtriangleup_s Y=\frac m2X_{,r}\ \ \mathrm{at}\ \ r=\frac m2\ .
\ee
In terms of $X,Y$ equations (\ref{12}) and  (\ref{23})  read
\be\label{28}
\bigtriangleup_s \tilde F=\frac{1}{\alpha}(z^2-1)\p_z(rX_{,r}+\frac 12\psi_0\bigtriangleup_s Y+(\psi_0-\frac{3m}{r\psi_0})Y)\ ,
\ee
\be\la{29}
\tilde P_I=\frac m2\int_{\frac m2}^{\infty}{\frac{dr}{r^2\psi_0} \int_{-1}^1(\alpha \tilde F_{,z}^{\ 2}- X\bigtriangleup_s Y)dz}+\frac 14\int_{-1}^1{(X^{2})_hdz}\ ,
\ee
where function $\tilde F$ is related to $F$ by 
\be
\tilde F=\psi_0^{-3}F\ .
\ee

Let us expand functions $X$ and $Y$ into the Legendre polynomials $P_n$,
\be\la{32}
X=\Sigma_{n=0} X_nP_n\ ,\ \ Y=\Sigma_{n=1}Y_nP_n
\ee
(note that (\ref{26}) admits $Y_0=0$). Condition (\ref{27}) yields
\be\la{32a}
X_{0,r}=0\ \  \mathrm{at}\ \  r=\frac m2\ ,
\ee
\be\la{32b}
Y_n=\frac {m}{2(2N+1)}X_{n,r}\ \ \mathrm{at}\ \ r=\frac m2\ ,\ n\geq 1\ ,
\ee
where
\be\la{41a}
N=\frac 12n(n+1)\ .
\ee
The integrability condition of (\ref{28}), equivalent to  (\ref{22}), leads to relation
\be\label{30}
Y_1=\frac{r^2\psi_0}{3m}X_{1,r}\ ,
\ee
 compatible with (\ref{32b}).
From (\ref{28}) one obtains
\be\la{33}
\tilde F_{,z}=\frac{1}{\alpha}\Sigma_{n=2} (rX_{n,r}-\beta_nY_n)\tilde P_n\ ,
\ee
where
\be\la{40}
\beta_n= (N-1)\psi_0+\frac{3m}{r\psi_0}
\ee
and
\be\la{34}
\tilde P_n=\p_z(\bigtriangleup_s^{-1}((z^2-1)\p_zP_n))=\frac{1}{n+2}(nP_n-\frac{2}{n-1}P_{n-1,z})\ ,\ n\geq 2\ .
\ee
Polynomials $\tilde P_n$ are orthogonal,
\be\la{35}
\int_{-1}^1{\tilde P_k\tilde P_ndz}=c_n\delta_{kn}\ ,
\ee
where
\be\la{36}
c_n=\frac{2n(n+1)}{(n-1)(n+2)(2n+1)}\ .
\ee

From (\ref{29}) and (\ref{33}) it follows that
\be\la{37}
\tilde P_I= \frac {2m}{3} I_1+\frac m2\Sigma_{n=2}c_nI_n+\frac 12\Sigma_{n=0}\frac{X_{nh}^2}{2n+1}\ ,
\ee
where
\be\la{38}
I_1=\int_{\frac m2}^{\infty}{ X_1Y_1\frac{dr}{r^2\psi_0}}
\ee
and 
\be
\begin{aligned}\la{40a}
I_{n\geq 2}&=\int_{\frac m2}^{\infty}{\big[(rX_{n,r}-\beta_nY_n)^2+2(N-1)\alpha X_nY_n]\frac{dr}{\alpha r^2\psi_0} }\\
&=\int_{\frac m2}^{\infty}{\frac{\beta_n^2}{\alpha r^2\psi_0}(Y_n-\frac{r}{\beta_n}X_{n,r}+\frac{(N-1)\alpha}{\beta_n^2}X_n)^2dr}-\frac{N-1}{m(2N+1)}X_{nh}^2\ .
\end{aligned}
\ee
For a fixed value of $X_{nh}$ functional $I_{n\geq 2}$ takes its minimum    if 
\be\la{41}
Y_n=\frac{r}{\beta_n}X_{n,r}-\frac{(N-1)\alpha}{\beta_n^2}X_n
\ee
(note that (\ref{41}) is compatible with (\ref{32b})).
Hence
\be\la{40c}
I_{n}\geq
-\frac{N-1}{m(2N+1)}X_{nh}^2\ ,\ \ n\geq 2
\ee
and from (\ref{30}) and (\ref{38}) one obtains
\be\la{42}
I_1=\frac{1}{6m}(X_{1\infty}^2-X_{1h}^2)\ .
\ee
Substituting (\ref{40c}) and (\ref{42})  into (\ref{37}) leads to 
\be\la{43}
\tilde P_I\geq \frac 19 X_{1\infty}^2+\frac 12\Sigma_{n=0}\frac{X_{nh}^2}{(2n+1)(n^2+n+1)}
\ee

To complete estimation of the r.h.s. of (\ref{32g}) we still need a harmonic function $\psi_1$, which satisfies the boundary condition
\be\la{47}
\p_r\psi_1+\frac 1m\psi_1= \frac{1}{m^2}X_h\ \ \mathrm{on}\ \  S_h\ 
\ee
 following from (\ref{6x}) and (\ref{46a}).
Since function $\psi_1$ vanishes at $\infty$, it is given by
\be\la{45}
\psi_1=\Sigma_0^{\infty}a_n(\frac{m}{2r})^{n+1}P_n\ .
\ee
Condition (\ref{47}) imply
\be\la{46}
a_n=-\frac{1}{m(2n+1)}X_{nh}\ .
\ee
Hence,
\be\la{47a}
\langle \psi_1^2\rangle_{h}=\frac{4\pi}{m^2}\Sigma_0^{\infty}\frac{1}{(2n+1)^3}X_{nh}^2\ .
\ee
Substituting (\ref{43}), (\ref{46a}) and (\ref{47a}) into (\ref{32g}) leads to 
\be\la{45}
P_I\geq P_J+\frac 19 X_{1\infty}^2+\Sigma_{2}^{\infty}\frac{(n-1)(n+2)X_{nh}^2}{2(n^2+n+1)(2n+1)^3}\ .
\ee
It follows from the ADM formula  that the linear momentum (directed along the symmetry axis) is given by
\be\la{48}
p=\frac 13 X_{1\infty}\ .
\ee
As in \cite{kt}, the angular momentum $J$ enters into a structure of the function $\omega$ coming from regularity assumption about $A_{ij}$,
\be\la{50}
\omega=f(1-z^2)^2+J(z^3-3z)+c\ ,
\ee
hence
\be\la{52}
P_J\geq \frac{J^2}{4m^2}\ .
\ee
Substituting (\ref{48}),  (\ref{52}) and (\ref{46a}) into  (\ref{45})  yields
\be\la{53a}
E^2-p^2\geq \frac{|S_h|}{16\pi}+\frac{4 \pi}{|S_h|}J^2+\Sigma_2^{\infty}\frac{(n-1)(n+2)}{2(n^2+n+1)(2n+1)^3}X_{nh}^2\ .
\ee
It follows from (\ref{8}), (\ref{9}) and (\ref{26}) that on $S_h$ one has
\be\la{46a}
X_h=\frac{m^2}{32}(B_{rr}-\frac 23B\psi_0^6)_h\ .
\ee
Thanks to (\ref{46a}),  inequality (\ref{53a}) for maximal data ($B=0$) coincides with that in \cite{kt}. Inequality (\ref{53a})  is saturated if $f=0$ and (\ref{41}) is satisfied.

For the purpose of this paper let us define generic  data in the following way.
\begin{definition}
We call data generic if at least one of the following conditions is not satisfied:
\begin{enumerate}
 \item
 The leading terms (of order $\epsilon$) in  $A_{\varphi\theta}$ and $A_{\varphi r}$ are given by $B_{\varphi\theta}=0$ and $B_{\varphi r}=3Jr^{-2}\sin^2{\theta}$.
 \item
On the MOTS $(B_{rr}-\frac 23B\psi_0^6)_h$ is linear in $z$.
 \item All moments $Y_n$ with $n\geq 2$ are defined by $X_n$ according to (\ref{41}).
 \end{enumerate}
  \end{definition}

Condition 1 corresponds to $f=0$ in $\omega$. Condition 2 is equivalent to $X_n=0$ for $n\geq 2$. Concerning Condition 3, let us note that equation (\ref{41}) is also satisfied if $n=1$ (see (\ref{30})). Since $-n(n+1)$ are eigenvalues of $\bigtriangleup_s$, equation (\ref{41}) multiplied by $\beta^2_n$ is equivalent to
\be\la{54}
\psi_0^2(\bigtriangleup_s+2-\frac{6m}{r\psi_0^2})^2Y=-2r\psi_0(\bigtriangleup_s+2-\frac{6m}{r\psi_0^2})X_{,r}+2\alpha(\bigtriangleup_s+2)X+f(r)\ ,
\ee
where  function $f$ can be adjusted to $X_0$. Action of $\bigtriangleup_s$ on (\ref{54}) yields the following  equation for original leading terms in $K'_{ij}$,

\be\la{55}
\begin{split}
&(\bigtriangleup_s+2-\frac{6m}{r\psi_0^2})[(B_{\theta\theta}+\frac 12r^2B_{rr})\sin^2{\theta}]_{,zz} \\
&=(\bigtriangleup_s+2)[\frac{\alpha}{\psi_0}(rB_{r\theta}\sin{\theta})_{,z}-\frac 12r^2\bigtriangleup_s (B_{rr}-\frac 23 B\psi_0^6)].
\end{split}
\ee
Thus, condition 3 is equivalent to the additional constraint (\ref{55}) on initial data.

\null

   \section{Summary}
   Let us formulate the main result of this paper  in a similar way as in \cite{kt}.
 
\begin{thm}
  Let $S=R^3\backslash B(0,\frac m2)$ be an initial surface bounded by the sphere $S_h$ with radius $m/2$ and $g_{ij}$ be flat metric on $S$.
  Assume that constraints (\ref{4x})-(\ref{6x}) admit axially symmetric asymptotically flat solution  depending on a parameter $\epsilon$ in agreement with expansions (\ref{7x})-(\ref{8x}).
 
 Then,  in generic case (see Definition 3.1), initial data (\ref{3x}) satisfy the sharp Penrose inequality (\ref{8k}) up to the second order in $\epsilon$.
 \end{thm}
This theorem generalizes partially that in \cite{kt}, but now an analysis of nongeneric case is much more difficult. Note that data which satisfy conditions 1-3 of Definition 3.1 depend on one free function $X$. Moreover, in order to include terms $\epsilon^3$ and $\epsilon^4$ in $P_I$ we have to solve (\ref{4x}) up to terms proportional to $\epsilon^2$. These technical difficulties are rather discouraging. 

Concerning further generalizations, admitting data depending on the azimuthal angle $\varphi$ seems feasible, but a change of shape of the MOTS or admitting metrics which are not  conformally flat would require a new approach not based on expansions into the Legendre polynomials.

\end{document}